\documentclass[conference]{IEEEtran}

 \newcommand{\myfs}{\vspace{-0.0cm}}

\def\longversion{1}

\usepackage{amsmath,amssymb,multicol,wrapfig}
\usepackage{nicefrac}

\newtheorem{theorem}{Theorem}
\usepackage{url}
\usepackage{algorithm,algorithmic}
\usepackage[font=footnotesize]{subfig}
\usepackage{booktabs,multirow}

\newcommand{\exptime}{\textsc{exptime}}

\newtheorem{defn}{Definition}

\usepackage[noshell]{pdftricks}
\usepackage{pdftricks,psgo}
\begin{psinputs}
  \usepackage{pstricks}
  \usepackage{psgo}
\end{psinputs}

\newcommand{\kyu}{\text{k}}
\newcommand{\dan}{\text{D}}

\newcommand{\OMIT}[1]{}

\usepackage{acronym}
\acrodef{MCTS}{Monte Carlo Tree Search}
\acrodef{PLC}{Playing-level Complexity}
 
\title{Depth, balancing, and limits of the Elo model}

\author{
	\IEEEauthorblockN{
 Marie-Liesse Cauwet, 
 Olivier Teytaud 
}
	\IEEEauthorblockA{TAO, Inria, Univ. Paris-Sud, UMR CNRS 8623\\                
	Email: {\em \{firstname.lastname\}@inria.fr}\\
	Paris, France }\\
        \IEEEauthorblockN{
 Tristan Cazenave, 
 Abdallah Saffidine 
 }
	\IEEEauthorblockA{
	LAMSADE, Universit\'e Paris-Dauphine,\\ Paris, France\\
	}
	\and
        \IEEEauthorblockN{
 Hua-Min Liang,
 Shi-Jim Yen
}                                                            
	\IEEEauthorblockA{
	Computer Science and Information Engineering,\\ National Dong Hwa University,\\ Hualien, Taiwan
	}\\
        \IEEEauthorblockN{
 Hung-Hsuan Lin,
 I-Chen Wu
}
	\IEEEauthorblockA{
	Computer Science and Information Engineering,\\ National Chao Tong University,\\ Hsinchu, Taiwan
	}
}

\usepackage{epstopdf}	 

\begin{document}
\maketitle
\begin{abstract}
Much work has been devoted to the computational complexity of games. However, they are not necessarily relevant for estimating the complexity in human terms. Therefore, human-centered measures have been proposed, e.g. the depth.
This paper discusses the {\em{depth}} of various games, extends it to a continuous measure.
We provide new {\em{depth}} results and present tool (given-first-move, pie rule, size extension) for increasing it.
We also use these measures for analyzing games and opening moves in Y, NoGo, Killall Go, and the effect of pie rules.
\end{abstract}

\section{Introduction}

Combinatorial or computational measures of complexity are widely used for games, specific parts of games, or families of games~\cite{kasai79,HearnDemaine2009,Viglietta2012,BonnetJamainSaffidine2013IJCAI}.
Nevertheless, they are not always relevant for comparing the complexity of games from a human point of view{\color{black}: 
\begin{itemize}
\item we cannot compare various board sizes of a same game with complexity classes P, NP, PSPACE, EXP, \dots because they are parametrized by the board size; and for some games (e.g. Chess) the principle of considering an arbitrary board size does not make any sense.
\item state space complexity is not always clearly defined (for partially observable games), and does not indicate if a game is hard for computers or requires a long learning.
\end{itemize}
}
In Section \ref{humcompl}, we investigate another perspective that aims at better reflecting a human aspect on the {\em{depth}} of games.  Section \ref{defidepth} defines the {\em{depth}} of games and the related \ac{PLC} is described in Section \ref{plc}. In Section \ref{somegames} we review the complexity of various games. In Section \ref{boardsize} we see the impact of board size on complexity. In Section \ref{govskag}, we compare the {\em{depth}} of Killall-Go to the {\em{depth}} of Go.

In Section \ref{th}, we analyze how various rules concerning the first move impact the {\em{depth}} and PLC. We focus on the pie rule (PR), which is widely used for making games more challenging, more balanced. 

Then, we switch to experimental works in Section \ref{prxp}.  We study in Section~\ref{sec:altering} how the PR alters the \ac{PLC} of existing games such as NoGo, Y, and Chinese Dark Chess (Fig. \ref{nogoycdc}). Section \ref{dkg} then analyzes the {\em{depth}} and \ac{PLC} of Killall-Go, in particular when using PR. 
{\begin{figure}[t]
	\centering\myfs
\subfloat[\label{fig:Nogo} NoGo]{
	\begin{minipage}{.21\textwidth}
	\includegraphics[width=.99\textwidth]{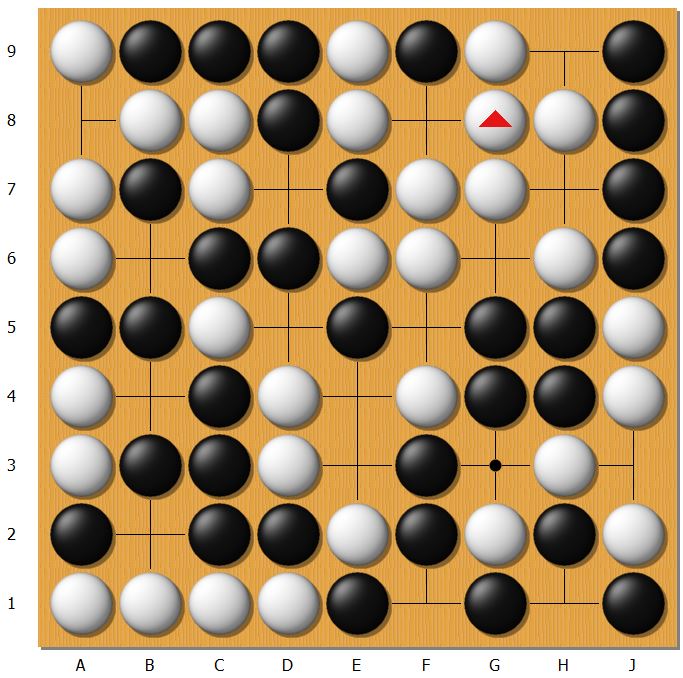}
	\end{minipage}
}\hfill
\subfloat[\label{fig:Y} Y]{
	\begin{minipage}{.15\textwidth}
		\includegraphics[width=\textwidth]{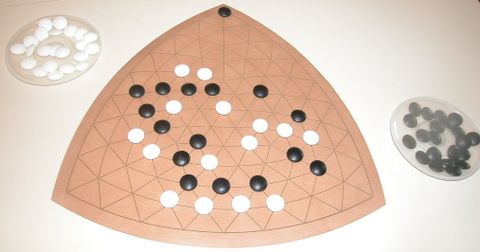}\\
		\includegraphics[width=.6\textwidth]{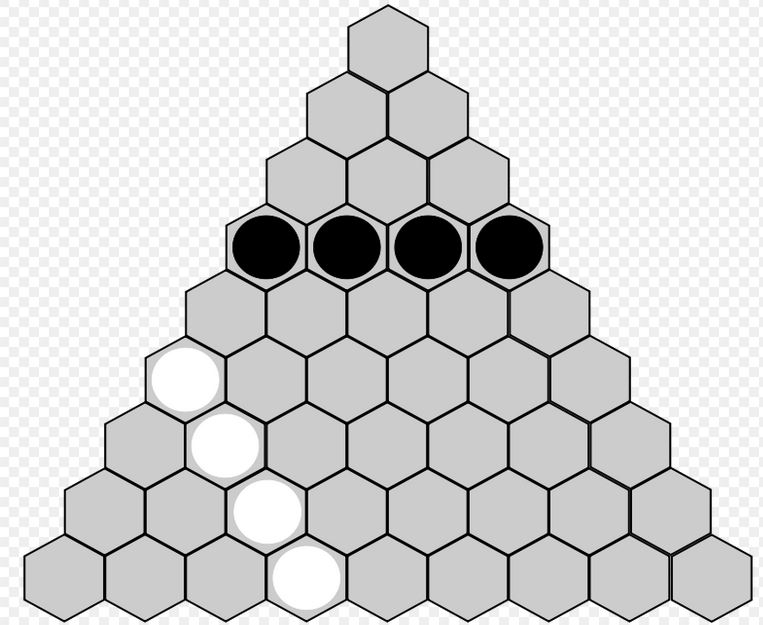}\\
	\end{minipage}
}\hfill
\subfloat[\label{fig:CDC} Chinese Dark Chess]{
	\begin{minipage}{.28\textwidth}
	\includegraphics[width=.99\textwidth]{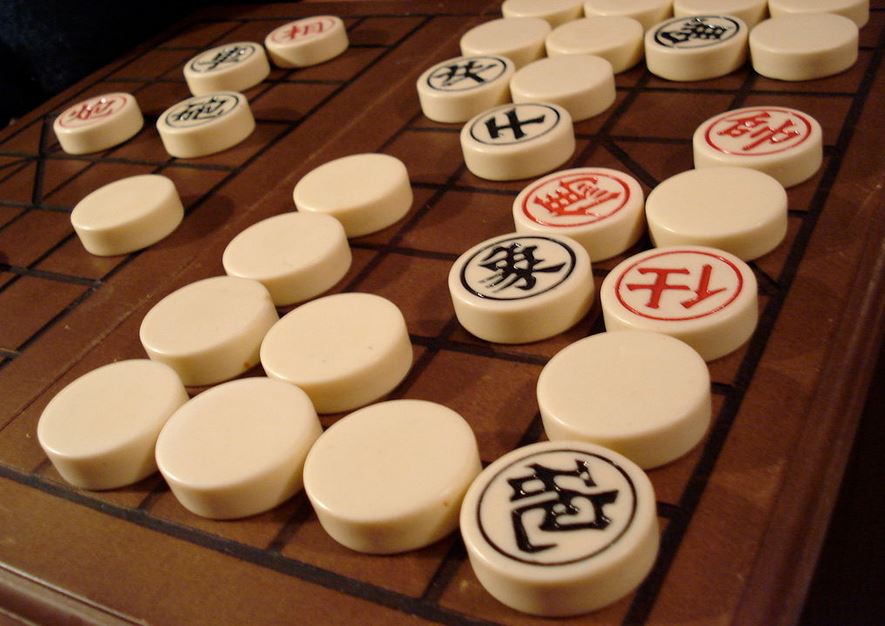}
	\end{minipage}
}
	\caption{\small \label{nogoycdc}The NoGo, Y (2 variants) and Chinese Dark Chess games respectively. NoGo has a gameplay similar to Go, but capturing is forbidden which completely changes the strategy (this game is a win by BobNogo against NdhuNogo, July 2012 tournament). Y is a connection game: players play in turn and the first player who connects 3 sides wins (corners are connected to both sides). Chinese Dark Chess is related to Chinese Chess, but with hidden pieces; a special move consists in turning a face-down piece into face-up. Sources: Wikipedia and Kgs.}
\end{figure}}
In all the paper, a rational choice means a choice which maximizes the success rate. This is the behavior of a player willing to win and having access to all possible information. In many cases below, we will assume that the opening moves and the pie rule choices (to swap or not to swap) are rational. In all the paper, $\log$ denotes logarithm with basis $10$.

\section{Human-centered complexity measures}\label{humcompl}
In this section, we review some human-centered complexity measures for games: the {\em{depth}} and the playing level complexity.
\subsection{Definition of the {\em{depth}} of a game}\label{defidepth}
{{
\begin{defn}
Consider a game $G$, and a set $S$ of players.
Then the {\em{depth}} $\Delta$ of $G$ is the maximal size of a set $p_1,\dots,p_{\Delta}$ of players in $S$ such that for each $i$, $p_i$ wins with probability $\geq 60\%$ against $p_{i+1}$. 
\end{defn}
}}
This measure was used in particular in~\cite{depth}.
{{This definition is less than definitive: the {\em{depth}} of a game depends on whether we consider computer players, or just humans, and among them, any possible player, or only players playing in a ``reasonable'' manner.}}
Also, many stupid games can be very deep for this measure: for example, who was born first, or who has the largest bank account number, or who has the first name in alphabetic order. 
The {\em{depth}} depends on the set of players.
Let us consider an example showing that most games are extremely deep for this measure, if we consider the {\em{depth}} for all possible players. If, using the gameplay, players can exchange $b$ bits of information, then one can build $1,\dots,2^b$ players $p_1,\dots,p_{d}$, with $p_i$ winning almost surely against $p_{i+1}$, as follows:
\begin{itemize}
	\item First, $p_i$ starts by writing $i$ on $b$ bits of information (in Go, $p_i$ might encode this on the top of the board if he is black, and on the bottom of the board if he is white);
 	\item Second, if the opponent has written $j<i$, then $p_i$ resigns.
\end{itemize}
This ensures an exponential {\em{depth}} of Go on an $n\times n$ board (and much better than this exponential lower bound is possible), with very weak players only. So, with no restrictions on the set of players, {\em{depth}} makes no sense.

An advantage of depth, on the other hand, is that it can compare {{completely unrelated games (Section \ref{somegames}) or different board sizes for a same game (Section \ref{sec:impact-size}).}}

\subsection{Playing-level complexity}\label{plc}
We now define a new measure of game complexity, namely the playing level complexity (\ac{PLC}). 
{{
\begin{defn}
The \ac{PLC} of a game is the difference between the Elo rating of the strongest player and the Elo rating of a naive player for that game.
More precisely, we will here use, as a \ac{PLC}, the difference between the Elo rating of the best available player (in a given set) and the Elo rating of the worst available player (in the same set).
\end{defn}
}}

This implies that the \ac{PLC} of a game depends on the set of considered players, {{as for the classical {\em{depth}} of games}}. An additional requirement is that the notion of best player and the notion of worst player make sense - which is the case when the Elo model applies, but not all games verify, even approximately, the Elo model (see a summary on the Elo model in Section \ref{conc}). 

When the Elo model holds, the \ac{PLC} is also equal to the sum of the pairwise differences in Elo rating: if players have ratings $r_1<r_2<\dots<r_n$, then the playing level complexity is $plc=r_n-r_1=\sum_{i=1}^{n-1} (r_{i+1}-r_i)$. Applying the Elo classical formula, this means
\begin{small}
\begin{equation}
plc=-\sum_{i=1}^{n-1} 400\log(1/P_{i+1,i}-1),\label{iAmInAPeacefulPlace}
\end{equation}
\end{small}
where $P_{i+1,i}$ is the probability that player $i+1$ wins against player $i$. When the Elo model applies exactly, this is exactly the same as 
\begin{small}
\begin{equation}
	plc=-400\log(1/P(strongest\ wins\ vs\ weakest)-1).\label{plcelo}
\end{equation}
\end{small}
While both equations are equal in the Elo framework, they differ in our experiments and Eq. \ref{iAmInAPeacefulPlace} is closer to {\em{depth}} results.

Unlike the computational complexity and just as the state-space complexity, the \ac{PLC} allows to compare different {{board}} sizes for the same game.
We can also use the \ac{PLC} to compare variants that cannot be compared with the state-space complexity, such as {{different starting positions or various balancing rules, e.g. imposed first move or pie rules.}}

{{When the Elo model applies,}} the \ac{PLC} is related to the depth. Let us see why. Let us assume that:
(i) the Elo model applies to the game under consideration, 
(ii) there are infinitely many players, with players available for each level in the Elo range from a player $A$ (weakest) to a player $B$ (strongest).
{{
Then, the {\em{depth}} $\Delta$ of the game, for this set of players, is proportional to the \ac{PLC}:
\begin{small}
\begin{equation*}
	\Delta=1+\left\lfloor \frac{plc}{ -400\log(\nicefrac{1}{.6}-1)}\right\rfloor=\left\lfloor\frac{Elo(B)-Elo(A)}{ -400\log(\nicefrac{1}{.6}-1)}\right\rfloor.
\end{equation*} \end{small}
.6 comes from the 60\%. The ratio is $-400\log(\nicefrac{1}{.6}-1)\simeq .70.437$.
}}
Incidentally, we see that the \ac{PLC} extends the {\em{depth}} in the sense that non-integer ``depths'' can now be considered.
\begin{table}[t!]\small
	\centering
	\caption{\small {\bf{Left:}} {\color{black}lower bounds on the \ac{PLC} and {\em{depth}} for some widely played games. } For Leagues of Legends the results are based on the Elo rating at the end of season 2 (data from the website competitive.na.leagueoflegends.com).  {\bf{Right:}} Analysis of the PR for NoGo and Y.}
  \label{tab:games-complexity}
\subfloat[\label{tab:fatigue}\scriptsize The \ac{PLC} is obtained by difference between the maximum Elo and the Elo of a beginner. Depths are obtained from these \ac{PLC} for these games, and/or from \cite{depth}.]{\scriptsize
\begin{tabular}{lrr}
    \toprule
    Game & \ac{PLC} & D. \\
    \midrule
    Go $19\times 19$    &             & $\geq$ 40 \\
    Chess               & $2300$ & $\geq$ 16 \\
    Go $9\times 9$      & $2200$ & $\geq$ 14 \\
    Ch. Chess       &             & $\geq$ 14 \\ 
    Shogi               &             & $\geq$ 11 \\
    L.o. Legends       & $1650$ & $\geq 10$ \\
    \bottomrule
  \end{tabular}
\begin{tabular}{lrr}
    \toprule
    Game & \ac{PLC} & D. \\
    \midrule
    Checkers            &             & $\geq$ 8 \\
    Backgammon          &             & $\geq$ 4 \\
    Urban Rivals        & $550$  & $\geq 3$ \\     
    Magic the G.&              & $\geq$ 3 \\
    Poker               &             & $\geq$ 1 \\
    \bottomrule
  \end{tabular}
  }
    \hspace{0mm}
  \subfloat[\label{tab:nogo-y}\scriptsize \acl{PLC} for various variants of the games NoGo and Y. RDR, RDR+pr and others refer to rules for managing the first move; see Sections \ref{th} and \ref{prxp}.]{ \scriptsize
\begin{tabular}{l@{\quad}r@{\quad}r@{\quad}r}
    \toprule
    Game, & RDR & RDR+PR & $\max_i $\ \\
   size,  &  & & \ $(RDR+$\\
    \# players  &  &  &\  $+GFM_{i}$)\\
    \midrule
    NoGo$5$ (7p) & 186.2 & 185.5 & 207.6 \\
    NoGo$6$ (7p) & 263.2 & 277.3 & 357.8 \\
    NoGo$7$ (4p) & 305.1 & 336.6 & 351.0 \\
    \midrule
    Y$4$ (6p) & 885.0 & 910.2 & 906.8 \\
    Y$5$ (7p) & 946.9 & 1185.7 & 1127.9 \\
    \bottomrule
  \end{tabular}
}
\end{table}
\subsection{{{Depth and PLC}} of various games}\label{somegames}
According to estimates from Elo scales and Bill Robertie's results~\cite{depth}, 
Chess would have {\em{depth}} 16, Shogi more than 11, checkers 8, Chinese Chess 14~\cite{dchessnew}, Backgammon a4, Magic The Gathering  3 or 4 (using data at \url{http://www.wizards.com/}), Poker 1 or 2 (see posts by  Ashley Griffiths Oct. 27th, 2011, on the computer-Go mailing list). The Chinese Chess estimate is based on Chinese Chess player Xiu Innchuan, currently Elo 2683.

Go is the deepest - up to now, as we will see that Killall-Go might outperform it. Go is a fascinating game for the simplicity of its rules, as well as for its tactical and strategic complexity (it is one of the games which can be naturally extended to bigger boards in an \exptime{}-complete manner \cite{exptimego}); it leads naturally to many variants. 
\ifthenelse{\longversion=1}{S. Ravera posted the following evaluation of Go's {\em{depth}} on the French Go mailing-list: Beginners have 100 points; 9D players on the EGF database have 2900 points; a difference of 400 points lead to 75\% winning probability for players between 20K and 6K; a difference of 300 points lead to 75\% winning probability for players between 5K and 3K; a difference of 200 points lead to 75\% winning probability for players between 2K and 3D; a difference of 100 points lead to 75\% winning probability for players between 4D and 6D. With these statistics, he used the threshold 75\% instead of 60\%, and concludes to 13 classes of players, namely ($20\kyu < 16\kyu < 12\kyu < 8\kyu < 4\kyu < 1\kyu < 2\dan < 4\dan < 5\dan < 6\dan < 7\dan < 8\dan < 9\dan$).
Numbers are based on the EGF database.
As the Elo difference for 75\% is $\frac{\log(1/.75-1)}{\log(1/.6-1)}=2.7$ times bigger than the Elo difference for 60\%, this suggests 35 complexity levels. This is probably underestimated as classes are constrained to be Go classes ($1\dan$, $2\dan$, $3\dan$, \dots; and not floating point numbers such as $1.4\dan$, $2.3\dan$, \dots), so the 40 estimate is probably reasonnable.}{S. Raveira posted on the French Go mailing-list a similar estimate using the EGF database.}
Using Elo ratings from \url{http://senseis.xmp.net/?EloRating}, we got a few more levels, leading to 40, as suggested in Robertie's paper~\cite{depth}. 
Let us estimate the {\em{depth}} of $9\times 9$-Go for computer players. In $9\times 9$-Go, computers are comparable to the best human players. The Cgos $9\times 9$-Go server gives a Elo range of roughly 2200 between random and the best computer player in the all time ratings, which suggests a {\em{depth}} of 14.  This is for games with 5 minutes per side, i.e., very fast.
We summarize these {\em{depth}} results in Table~\ref{tab:games-complexity}. Games with relatively high depth are difficult for computers; but games at the bottom are not all that easy, as illustrated by Backgammon \cite{TDgammon} or Poker \cite{billingsphd}.

\subsection{Impact of the board size: NoGo, Y and Chinese Dark Chess}\label{boardsize}
\label{sec:impact-size}
We have seen that Go 19x19 is much deeper than Go 9x9. In this section, we check that this is also the case for NoGo, Y and Chinese Dark Chess. We check this for various rules (RDR, RDR+PR, RDR+GFM), which are introduced later (the details are irrelevant for the comparison here).
The results on NoGo (Fig. \ref{fig:Nogo}) and Y (Fig. \ref{fig:Y}) in Table~\ref{tab:nogo-y} are also indicative of the relationship between the \ac{PLC} and the board size.
We see that for all variants, the game of Y appears to have a larger \ac{PLC} on size 5 than on size 4, with the same set of players (\ac{MCTS} with the same numbers of simulations).
Similarly the PLC of NoGo increases from size $5 \times 5$ to size $6\times6$, also with the same set of players (\ac{MCTS} with the same numbers of simulations).
{\begin{figure*}[t]\centering\myfs
		\centering
	\hfill
	\includegraphics[width=.27\textwidth]{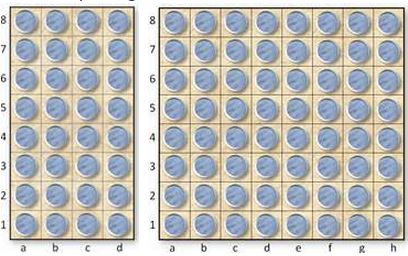}
	\hfill
	\includegraphics[width=.09\textwidth]{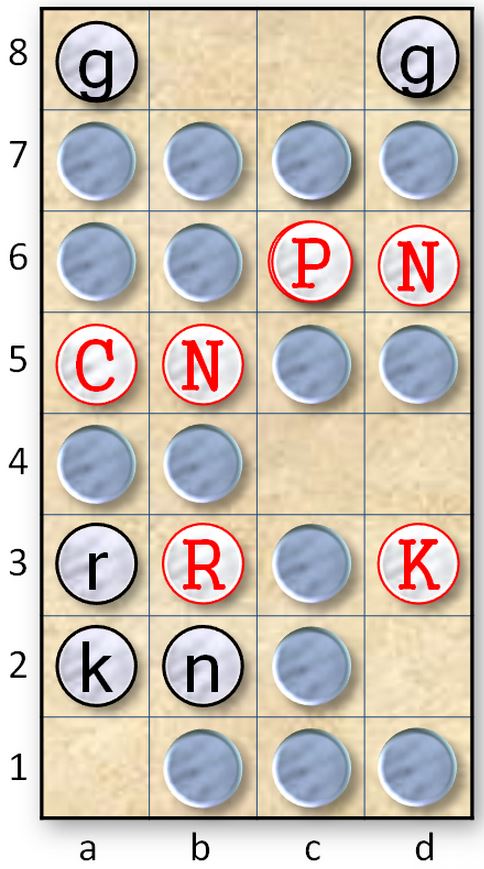}
	\hfill
\begin{minipage}{6cm}	\vspace*{-1.5cm}In the opening, both sides revealed a king. In the move sequences, 12. a5(C) b8-a8 13. a3(r) a3-b3 14. a5-a2, Red found Black's king could not move horizontally and tried to capture the king by means of revealing two pieces at squares a5 and a3. However, the pieces flipped at a5 was the red cannon which captured Black's king A2 in the subsequent moves. Estimating probabilities is crucial in CDC.  \end{minipage}
	\hfill\ 
	\caption{\label{cdc48}{\bf{Left:}} opening board in 4x8 CDC (left) and 8x8 (right): all pieces are face-down. {\bf{Right:}} Chinese Dark Chess game won by DarkKnight as the red player, in competition \cite{cdcdn}. 
}
\end{figure*}}
The game of Chinese Dark Chess (CDC) (Fig. \ref{fig:CDC}) is a stochastic perfect information game usually played on a $4 \times 8$ board.
It is possible to create a larger $8 \times 8$ board by putting two regular boards side by side and using two sets of pieces (Fig. \ref{cdc48}).
We present in Table~\ref{tab:cdc} the winning rates of our CDC program (an \ac{MCTS} program) against an Alpha-beta baseline player,
depending on the number of MCTS simulations.
We compute the Elo rating of the set of players assuming the baseline has a fix rating of 1000.
Using Eq.~\ref{plcelo} we derive that the \ac{PLC} for this set of players is 270.73 (depth 3.84) for $4 \times 8$ Chinese Dak Chess and 323.95 (depth 4.60) for $8 \times 8$ CDC.
We can therefore conclude that the $8\times8$ version has a bigger \ac{PLC}, at least for the given set of players.
{\begin{small}
\begin{table*}[t!]\small
  \centering
  \caption{\small Estimation of the \acs{PLC} for two variants of CDC using \acs{MCTS} programs (4x8 and 8x8).}
  \label{tab:cdc}
  \scriptsize
  \begin{tabular}{l*{10}{r}}
    \toprule
    \multirow{2}{18mm}{Thousands of Simulations} & \multicolumn{5}{c}{$4 \times 8$} & \multicolumn{5}{c}{$8 \times 8$} \\
    \cmidrule(lr){2-6} \cmidrule(lr){7-11}
     & Wins & Losses & Draws & Score & Elo & Wins & Losses & Draws & Score & Elo \\
    \midrule
    0.3 & 0.24 & 0.26 &  0.5 & 0.49 &  993.05 & 0.01 & 0.46 & 0.53 & 0.28 &  831.60 \\
    1   & 0.45 & 0.09 & 0.46 & 0.68 & 1130.94 & 0.06 & 0.25 & 0.69 & 0.41 &  933.18 \\
    2   & 0.52 & 0.07 & 0.41 & 0.73 & 1168.40 & 0.15 & 0.14 & 0.72 & 0.50 & 1003.44 \\
    3   & 0.61 & 0.05 & 0.34 & 0.78 & 1219.87 & 0.19 & 0.09 & 0.72 & 0.55 & 1034.68 \\
    5   & 0.63 & 0.05 & 0.33 & 0.79 & 1227.17 & 0.30 & 0.06 & 0.64 & 0.62 & 1085.04 \\
    25  & 0.68 & 0.04 & 0.28 & 0.82 & 1263.42 & 0.44 & 0.02 & 0.54 & 0.71 & 1155.54 \\
    \midrule
    \ac{PLC} &   &   &       &      & 270.73 &       &      &       &     & 323.95 \\
    \bottomrule
  \end{tabular}
\end{table*}\end{small}}
\subsection{Is Killall-Go deeper than Go?}\label{govskag}
Finally, we can also use the data to compare Killall-Go (KAG, discussed in details in Section \ref{dkg}) to Standard Go.
Fig. \ref{kgo99} (right) shows a huge improvement when comparing the \ac{PLC} of KAG to Go.
A detailed analysis shows that there is a regular increase in Elo difference (i.e. player $i+1$ is usually stronger than player $i$ in Fig. \ref{kgo99} (right)), but some specific values lead to a larger gap - in particular player $11$ wins very easily against player $10$ in KAG.
The difference between the \ac{PLC} of KAG and Go is minor in our other experiments with only 2 players, namely HappyGo 500 simulations per move vs HappyGo 400 simulations per move - but in the case with 15 players there is a huge difference between KAG and Go.
\section{Choice rules and depth: theory}\label{th}
There are various systems for choosing, in a game,
(i) the initial position (e.g. base placement in Batoo or handicap placement in Go or KAG), 
(ii) or the first move (e.g. in Twixt, Havannah, Hex or Y, where the first move is crucial),
(iii) or some parameter (e.g. komi in Go).
 These three aspects are very related; it is all about making a fair choice, so that the game is more challenging.
 (a) It might be fixed by the rules.
 (b) It might be fixed by the rules, and depend on the context; for example, the relevant komi in Go increases when players' strength increases, and the handicap in KAG decreases when the level increases.
 (c) It might be chosen by bidding~\cite{auctionkomi} or by pie rule.
The pie rule (PR \cite{pierule}) is a classical tool intended to compensate the advantage of the first player by allowing the second player to exchange (to swap) the roles. 
It is quite effective for making Hex \cite{hexPaper}, Havannah \cite{fabienHavannah} or Twixt fairer. 
We present the PR and show some disappointing properties of it in Section \ref{piebad}. In particular, the PR cannot systematically increase the winning rate of the strongest player. Various possible effects of PR are illustrated in Section \ref{exa}. Then, Section \ref{morethan2} shows that the PR can nonetheless have a positive impact; we will see that this is indeed the usual case.
We will refer to Eq. \ref{iAmInAPeacefulPlace} for all evaluations of \ac{PLC} in this section.

\subsection{Two players only: the PR does not increase the \ac{PLC} with two players only}\label{piebad}

Let us first show that the PR cannot systematically increasing the Elo difference between two players, compared to an ad hoc choice of the first move by an all-knowing referee.

Consider the following scenario where Black plays first.
There are two players named player 1 and player 2.
Without PR, the game starts with player being assigned a color at random, say player $i$ is assigned Black and player $3-i$ is assigned White.
The game proceeds with player $i$ playing the first Black move, then player $3-i$ plays White, \dots until the game is over.
With PR, the game starts with one player at random picking a first Black move.  The opponent then decides who is assigned Black, say $j$, and who is playing White, $3-j$ (swap decision).
The game proceeds from here with player $3-j$ playing White, then player $j$ playing Black, \dots until the game is over.
This means that with the PR, the second player has the possibility to change roles. For this reason, the first player should not play a very strong move; he should instead play a move which leads to position as balanced as possible.

Let us assume that player 1 is stronger than player 2. This means that from any position, if 1 plays against 2, one game as black and one game as white, the expected number of wins for player 1 (over these 2 games) is greater than, or equal to, the expected number of losses.

Let us define $p_i$ the winning probability for player 1 (against player 2) if playing as black for moves 3, 5, 7, 9, \dots (player 2 plays moves 2, 4, 6, 8, \dots) and first move is $i\in \{1,2,\dots,K\}$ (with $K$ the number of possible first moves). 
Let us define $q_i$ the winning probability for player 1 (against player 2) if playing as white (for moves 2, 4, 6,\dots; player 2 plays moves 3, 5, 7, \dots) and first move is $i\in \{1,2,\dots,K\}$. Let us simplify notations by assuming that there is no draw (otherwise, a draw will be half a win).  We are interested in finding the method for which the winning rate of player 1 (assumed to be stronger) is the greatest, so that the game has a greater game playing complexity. 

The three considered methodologies are:
\begin{itemize}
\item RDR: Randomly draw roles, no PR: the winning rate of player 1 depends on the first move as chosen by the player who plays first. Let us assume that player 1 optimizes the first move so that their winning rate is maximum.
\item RDR+PR: Randomly draw roles, and possibly change by PR.
\item RDR+GFM$_i$: (GFM stands for ``given first move'') Randomly draw roles, no PR, and the first move is given by the rules - it is the move with index $i$.
\end{itemize}

 Then the following equations give the success rate $w$ of player 1:
 (a) RDR+GFM$_i$: $w_{\textup{rdr+gfm},i}=(p_i+q_i)/2$ if we decide that the first move is $i$.
 (b) RDR: $w_{\textup{rdr}}=\frac12(\max_i p_i+\min_i q_i)$.
 (c) RDR+PR: $w_{\textup{rdr+pr}}=\frac12(\min_i \max(p_i,q_i)+\max_i \min(p_i,q_i)) $.
Then we claim:
\begin{theorem}
  \label{thm:pr-bounded}
  Consider the PR, when the first move and the PR choice (swap or no swap) are rational.
Then,
\begin{equation}
  w_{\textup{rdr+pr}}\leq \max_i w_{\textup{rdr+gfm},i}.\label{lm:eq}
\end{equation}
\end{theorem}
\ifthenelse{\longversion=1}{
	\def\versionicwu{
4. The proof for theorem seems hard to understand. Following is my proof. 
Let us assume three moves with winrates p1, p2, p3 for B and q1, q2, q3 for W. Without loss of generality, p1$>$q1, p2$>$q2, p3$>$q3. 
Let us assume the right hand side of (3) to be (p1+q1), which is the max for all pi+qi. 
Let us assume $min_i$ max(pi,qi) to be p2. 
Let us assume $max_i$ min(pi,qi) to be q3.
Assume by contradictory that (3) is wrong, that is, p2+q3 $>$ p1+q1. 
Then, since p3 $>$ p2, p3+q3 $>$ p2+q3 $>$ p1+q1, which is contradictory since p1+q1 is the max. 
If we draw a figure (with three moves), that would make it much easier to understand.

	}
\begin{proof}
Let us first show that 
{\begin{equation}\label{inter}
  \min_{i} \left(\frac{p_{i}+q_{i}}{2}+\frac{|p_{i}-q_{i}|}{2}\right) -\min_{i}\frac{|p_{i}-q_{i}|}{2}\leq\max_{i}\frac{p_{i}+q_{i}}{2}
\end{equation}}
For RDR+PR and RDR+GFM$_{i}$, exchanging $p_i$ and $q_i$, for some $i$, does not change the game. This is because:
\begin{itemize}
\item For RDR+GFM$_{i}$, the role (Black or White) will be randomly chosen, so that exchanging $p_i$ (probability of winning if I am Black) and $q_i$ (probability of winning if I am White) does not change the result.
\item For RDR+PR, the role (Black or White) is chosen by the second player, given the choice of $i$ by the first player. Therefore, exchanging $p_i$ and $q_i$ leads to the same choices, just exchanged between Black and White.
\end{itemize}

So we can assume, without loss of generality, that $p_i\leq q_i$ for any $i\in\{1,\dots,K\}$. Then, by denoting $q_{i_0}:=\min_{i} q_{i} =\min_{i} \left(\frac{p_{i}+q_{i}}{2}+\frac{|p_{i}-q_{i}|}{2}\right)$ and $\frac{q_{i_1}-p_{i_1}}{2}:=\min_{i}\frac{q_{i}-p_{i}}{2}= \min_{i}\frac{|p_{i}-q_{i}|}{2}$, we get:
{
$$\min_{i} \left(\frac{p_{i}+q_{i}}{2}+\frac{|p_{i}-q_{i}|}{2}\right) -\min_{i}\frac{|p_{i}-q_{i}|}{2} $$
$$=q_{i_0}-\frac{q_{i_1}-p_{i_1}}{2}\leq \frac{q_{i_1}+p_{i_1}}{2}\leq \max_{i}\frac{p_{i}+q_{i}}{2}.$$}
We now prove Eq. \ref{lm:eq}, using the classical equalities $\max(a,b)=\frac{a+b}{2}+\frac{|a-b|}{2}$ and $\min(a,b)=\frac{a+b}{2}-\frac{|a-b|}{2}$ for any $(a,b)\in \mathbb{R}^{2}$:\\
{$w_{rdr+pr}=\frac12(\min_i \max(p_i,q_i)+\max_i \min(p_i,q_i))$
$=\frac12(\min_i [\frac{p_{i}+q_{i}}{2}+\frac{|p_{i}-q_{i}|}{2}]+\underbrace{\max_i \left[\frac{p_{i}+q_{i}}{2}-\frac{|p_{i}-q_{i}|}{2}\right]}_{\leq \max_{i}\frac{p_{i}+q_{i}}{2}-\min_{i}\frac{|p_{i}-q_{i}|}{2}})$;}
then, by using Eq. \ref{inter}, we get $w_{rdr+pr}\leq \max_{i} \frac{p_{i}+q_{i}}{2}$, which is the expected result.
\end{proof}
}{The detailed proof is reported to \cite{longversion}.}
{\bf{Remark:}} The assumption in Theorem~\ref{thm:pr-bounded} means that $p_i$ and $q_i$ are known by both players.  We also note that in some cases, the inequality is strict.
Importantly, we here compared the result to the best (in terms of \ac{PLC}) possible choice for the initial move. This is the main limitation; the PR does not perform better than an omniscient referee who would choose the initial move for making the game as deep as possible specifically for these two players. 
In the following, $\max_i(RDR+GFM_i)$ denotes the best possible choice for the initial move, i.e. the move that give $ \max_i w_{\textup{rdr+gfm},i}$.

\subsection{General case: examples and counter-examples of impact of the PR on \ac{PLC}}\label{exa}
Two opposite behaviors might happen.
(a) Table~\ref{tab:jefatigue} presents an example showing that the PR can indeed decrease the success rate of the strongest player, compared to a choice of the 1st move by the 1st player. In this case,
$w_{\textup{rdr}}>w_{\textup{rdr+pr}}.$
(b) Table~\ref{tab:polalacafatigue} presents an example showing that the PR can greatly increase the success rate of the strongest player, compared to RDR. In this case,
$w_{\textup{rdr+pr}}> w_{\textup{rdr}}.$
\subsection{An artificial example in which the \ac{PLC} is increased by the PR, compared to any fixed move}\label{piegood}\label{morethan2}
We have seen above (Eq. \ref{lm:eq}) that, when only two players are considered, 
$w_{\textup{rdr+pr}}\leq \max_i w_{\textup{rdr+gfm},i}.$
Moreover this inequality might be strict.
In table \ref{tab:ex1}, we present an artificial game (with more than 2 players) in which
$w_{\textup{rdr+pr}}> \max_i w_{\textup{rdr+gfm},i}.$
\begin{table}[t!]
  \centering
  \caption{\small \label{tab:ex} Artificial games to illustrate possible effect of the PR. Each number in the table is the probability of winning for the strongest player, depending on who plays black and what the first move is.}
  \subfloat[\label{tab:jefatigue} The PR decreases the success rate of the strongest player from 95\% to 90\%.]{\scriptsize
    \begin{tabular}{ccc}
      \toprule
      & \multicolumn{2}{c}{Score of 1 vs.~2} \\
      \cmidrule(lr){2-3}
	& as B & as W \\
      \midrule
      move A &  100\% & 90\%\\
      move B &  50\% & 90\% \\
      \bottomrule
    \end{tabular}}
    \hspace{2mm}
  \subfloat[\label{tab:polalacafatigue} B is a very strong move. PR increases the success rate of the strongest player from 50.005\% to 75\%.]{\scriptsize
    \begin{tabular}{ccc}
      \toprule
      & \multicolumn{2}{c}{Score of 1 vs.~2} \\
      \cmidrule(lr){2-3}
	& as B & as W \\
      \midrule
      move A & 100\% &  50\%\\
      move B & 100\% & 0.01\% \\
      \bottomrule
    \end{tabular}
  }
    \hspace{2mm}
  \subfloat[\label{tab:ex1}PR works better than any fixed move assuming a set of 3 players.]{\scriptsize
    \begin{tabular}{ccccc}
      \toprule
      & \multicolumn{2}{c}{Score of 1 vs.~2} & \multicolumn{2}{c}{Score of 2 vs.~3} \\
      \cmidrule(lr){2-3} \cmidrule(lr){4-5}
      & as B & as W & as B & as W \\
      \midrule
      move A &    0.96 & 0.99 &   0.37 & 0.67\\
      move B &    0.71 & 0.95 & 0.68 & 0.94\\
      \bottomrule
    \end{tabular}}
  \hspace{2mm}
\end{table}
By referring to Table \ref{tab:ex1}, if the first move is fixed at A, then $Elo_3-Elo_2=13.905$, $Elo_2-Elo_1=636.426$.
If it is fixed at B, then $Elo_3-Elo_2=251.893$, $Elo_2-Elo_1=275.452$.
In both cases, the Elo range, i.e. the \ac{PLC} for those 3 players (Eq. \ref{iAmInAPeacefulPlace}), is less than 651. With PR (random choice of initial colors, the second player to play can switch), if each player plays the optimal strategy, then 
$Elo_3-Elo_2=126.97$ and $Elo_2-Elo_1=530.72$.
The \ac{PLC} for those 3 players is therefore more than $657$.
Therefore, with two players, PR can be successful in terms of {\em{depth}} only against RDR, and not against $\max_i(RDR+GFM_i)$. However, with at least $3$ players, depending on the game, PR can be or not successful in terms of {\em{depth}} both compared to RDR and compared to $\max_i(RDR+GFM_i)$.
\section{Applying the Pie Rule: real games}\label{prxp}
We study the impact of the PR on NoGo and Y (Section \ref{sec:altering}) and on the handicap choice in KAG (Section \ref{dkg}). The short version is that the PR increases the {\em{depth}} in particular when a wide range of players is considered.
\subsection{Rules for choosing the first move: NoGo and Y}
\label{sec:altering}
We now switch to experimental works on the choice of the first move, in particular for the PR. We study the impact of the PR in NoGo and Y.
We first make the assumption that each player plays the PR perfectly and chooses the initial move perfectly in RDR (they know the success rates corresponding to a given first move), though later we will check the impact of this assumption by checking its impact on the \ac{PLC}.
We experimented (i) RDR+PR (ii) RDR (iii) RDR+GFM$_{i}$ with the best choice of $i$ and (iv) RDR+GFM$_{i}$ with the worst choice of $i$.
The \ac{PLC} is evaluated by Eq.~\ref{iAmInAPeacefulPlace}.
The pool of players consists of \ac{MCTS} players with varying number of simulations per move.

{\bf{The NoGo game}} (Fig. \ref{fig:Nogo}), designed for the BIRS seminar, has the same gameplay as Go but the goal of the game is different. In NoGo captures and suicides are illegal and the first player with no legal move loses the game.
The players are \ac{MCTS} with 125, 250, 500, 1000, 2000, 4000, 8000 simulations per move in size $5\times5$ and $6\times6$, and 125, 250, 500, 1000 simulations per move in size $7\times7$.
We designed 6 opening moves for $5\times5$, 6 opening moves for $6\times6$, 10 opening moves for $7\times7$.
The experimental results are displayed in Table~\ref{tab:nogo-y}.
We also evaluated the \ac{PLC} of $5\times 5$ by Eq. \ref{iAmInAPeacefulPlace} without using the rational first move assumption (i.e. the first move is now chosen by the first player); we got 
{
\begin{eqnarray}
\mbox{\ac{PLC}}_{nogo5x5}=240.32, \label{nogo1}\\
\mbox{\ac{PLC}}_{nogo6x6}=278.58,\\
\mbox{\ac{PLC}}_{nogo7x7}=429.71\label{nogo3}
\end{eqnarray}
}
which are all significantly higher than their ``rational opening'' counterparts and show that opening moves are not rational - and that the ability to decide the first move contributes significantly to the {\em{depth}} of the game.
We estimated the winrates used in the \ac{PLC} computations over 500 games for each setting.

{\bf{The game of Y}} is a connection board game invented by Shannon in the 50s (Fig. \ref{fig:Y}).

We test all the possible opening moves, and the players use $4^0$, $4^1$,\dots,$4^m$ simulations per move, with $m=5$ for board size 4 and $m=6$ for board size 5.
We estimated the winrates used in the \ac{PLC} computations over 20,000 games for each setting.
The experimental results are displayed in Table~\ref{tab:nogo-y}.
In NoGo, we see that, in 6x6 and 7x7, PR improved the \ac{PLC} compared to RDR, but not always when compared to $\max_i(RDR+GFM_i)$ - though it was successful for both cases with at least 7 players, which is consistent with the fact that with only 2 players PR cannot beat $\max_i(RDR+GFM_i)$ in terms of {\em{depth}} (Theorem \ref{thm:pr-bounded}).
In the case of the game of Y, the PR is a clear success.
Using the Pie Rule in Y extends the {\em{depth}} more than any other rule.

\subsection{Rules for choosing the handicap: Killall-Go}\label{dkg}
In KAG, the gameplay is the same as in Go, and the rules for deciding the winner are the same; but the komi (number of points given to black as a territory bonus for deciding the winner) is such that White wins if and only if he has at least one stone alive at the end. 
Deciding the handicap placement and the fair handicap is non trivial. For example, most people consider that living is hard for White in 9x9 with handicap 4; but Fig. \ref{kgo99} shows an opening which might be a solution for White to live.
{
{\begin{figure}[t]\centering\myfs
	\center
	\begin{minipage}{.4\textwidth}
	\includegraphics[width=.75\textwidth]{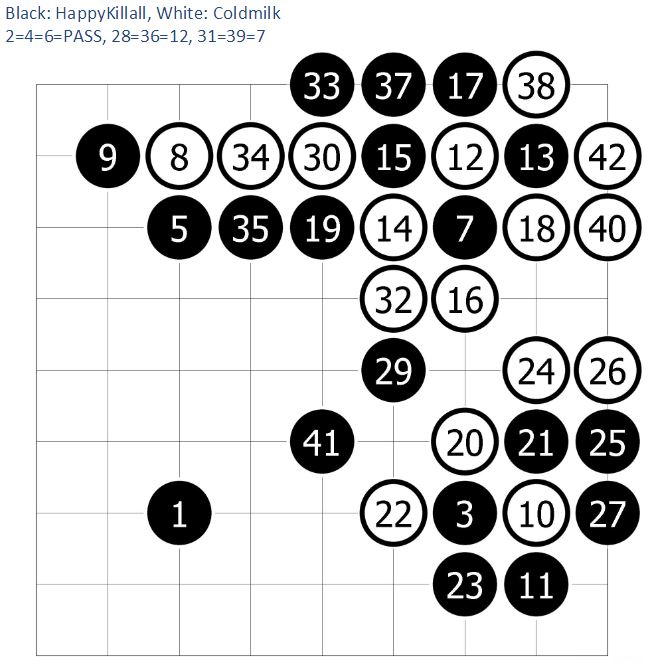}
	\end{minipage}
	\begin{minipage}{.55\textwidth}\scriptsize
\begin{tabular}{c|c|c|c|c}
	\hline
	 KAG ; nb        & KAG        & KAG        &  \\	
	of players& \ac{PLC} with & \ac{PLC} with & \ac{PLC} with & \ac{PLC}  \\
	$N$ & 4 stones hand. & best hand. & pie rule & of Go \\
	\hline
5 & 275 & 470 & 434 & 253\\
6 & 554 & 554 & 588 & 268\\
7 & 745 & 745 & 779 & 283\\
8 & 774 & 774 & 1011 & 292\\
9 & 774 & 999 & 1290 & 311\\
10 & 774 & 999 & 1290 & 328\\
11 & 1191 & 1191 & 1570 & 329\\
12 & 1382 & 1382 & 1690 & 329\\
13 & 1382 & 1382 & 1690 & 333\\
14 & 1382 & 1382 & 1690 & 347\\
15 & 1382 & 1382 & 1690 & 349\\
	\hline
\end{tabular}
\end{minipage}
	\caption{Top: \label{kgo99} KAG with handicap 4 in 9x9. White wins (i.e. White lives), thanks to an elegant opening. Bottom:  Comparison between the {\em{depth}} of KAG (columns 2 to 4) and Go (column 5) for different numbers $N$ of GnuGo players (first column). Row $N$ corresponds to the {\em{depth}} for the set of players $(1,2,3,\dots,N)$, where player $i$ has $50\times i$ simulations per move.}
\end{figure}}
}
The Black player is also known as \emph{Killer} and White is known as \emph{Defender}.
In KAG, the handicap is naturally much larger than in a standard Go game, in order to compensate the special komi.
For example, 17 handicap stones is common in $19\times 19$ KAG.
We use the PR for choosing the number of handicap stones, with, then, free handicap positioning (Black chooses the position). 
The handicap can be 2, 3, 4 or 5; we consider 9x9 Kill-all Go.
We consider $N$ players, which are GnuGo/\ac{MCTS} with $50,100,150,\dots,50\times N$ simulations per move respectively. Besides the best handicap choice (in the sense: the one which maximises the depth) and the pie rule, we consider the 4 stones handicap, which is the most reasonable for strong players.
We get results as presented in Fig. \ref{kgo99}, right.
Except for very weak players the handicap which maximizes the \ac{PLC} between players is 4. We also see that the pie rule, as soon as the number of players is large enough, increases the \ac{PLC} compared to fixing once and for all the \ac{PLC}. The surprising gap between the \ac{PLC} of KAG and the \ac{PLC} of Go is discussed in Section \ref{govskag} - this gap is important but highly depends on the set of considered players.
\def\notclearenough{
\subsection{Deciding the handicap placement}

The previous section has shown how the pie rule is effective for choosing the number of handicap stones. We now check whether it is effective for choosing the detailed handicap placement. This means that the first player chooses the handicap positionning, and the second player chooses who is Black.
Then, we will check which position the referee should choose for making the game as deep as possible.

\subsubsection{Selecting handicap placement by PR}
Let us consider a free handicap stones placement in our KAG.
We could consider the handicap placement as the first move of the game, but the number of legal handicap stones placements is daunting.
As a result, how to artificially select a good handicap stones placement remains an open research problem.
Conversely, the handicap stones placement problem can be seen as a challenging benchmark for AI techniques.
In this work, we adopt the following approach.
For each board size we are interested in, we create a set of plausible handicap placements which are then further studied from a computational perspective.
\ifthenelse{\longversion=1}{
The 12 3-stone handicap placements we propose for board size $7 \times 7$ are displayed in Fig.~\ref{fig:handicap-7x7}.
Figure~\ref{fig:handicap-9x9} shows 10 4-stone placements for $9 \times 9$, and Fig.~\ref{fig:handicap-13x13} shows 4 10-stone placements for $13 \times 13$.}{The 12 3-stone handicap placements we propose for board size $7\times 7$, the 10 4-stone placements we propose for $9\times 9$, and the four 10-stone placements we propose for $13\times 13$ are defined in \cite{longversion}.} The PR will be used for selecting the handicap placement among these possibilities.
\ifthenelse{\longversion=1}{
{\begin{small}	\begin{figure}[t]\centering\myfs
  \centering
  \subfloat[\label{fig:handicap-77-3a}]{\begin{minipage}{0.25\columnwidth}\input{variantsGo/handicap-77-3a}\end{minipage}}
  \hfill
  \subfloat[\label{fig:handicap-77-3b}]{\begin{minipage}{0.25\columnwidth}\input{variantsGo/handicap-77-3b}\end{minipage}}
  \hfill
  \subfloat[\label{fig:handicap-77-3c}]{\begin{minipage}{0.25\columnwidth}\input{variantsGo/handicap-77-3c}\end{minipage}}
  \hfill
  \subfloat[\label{fig:handicap-77-3d}]{\begin{minipage}{0.25\columnwidth}\input{variantsGo/handicap-77-3d}\end{minipage}}
  \hfill
  \subfloat[\label{fig:handicap-77-3e}]{\begin{minipage}{0.25\columnwidth}\input{variantsGo/handicap-77-3e}\end{minipage}}
  \hfill
  \subfloat[\label{fig:handicap-77-3f}]{\begin{minipage}{0.25\columnwidth}\input{variantsGo/handicap-77-3f}\end{minipage}}
  \hfill
  \subfloat[\label{fig:handicap-77-3g}]{\begin{minipage}{0.25\columnwidth}\input{variantsGo/handicap-77-3g}\end{minipage}}
  \hfill
  \subfloat[\label{fig:handicap-77-3h}]{\begin{minipage}{0.25\columnwidth}\input{variantsGo/handicap-77-3h}\end{minipage}}
  \hfill
  \subfloat[\label{fig:handicap-77-3i}]{\begin{minipage}{0.25\columnwidth}\input{variantsGo/handicap-77-3i}\end{minipage}}
  \hfill
  \subfloat[\label{fig:handicap-77-3j}]{\begin{minipage}{0.25\columnwidth}\input{variantsGo/handicap-77-3j}\end{minipage}}
  \hfill
  \subfloat[\label{fig:handicap-77-3k}]{\begin{minipage}{0.25\columnwidth}\input{variantsGo/handicap-77-3k}\end{minipage}}
  \hfill
  \subfloat[\label{fig:handicap-77-3l}]{\begin{minipage}{0.25\columnwidth}\input{variantsGo/handicap-77-3l}\end{minipage}}
  \caption{\small \label{fig:handicap-7x7} Handicap positions used for $7\times7$ KAG.}
\end{figure}\end{small}}
{\begin{small}\begin{figure}[t]\centering\myfs
  \centering
  \subfloat[\label{fig:handicap-99-4a}]{\begin{minipage}{0.25\columnwidth}\input{variantsGo/handicap-99-4a}\end{minipage}}
  \hfill
  \subfloat[\label{fig:handicap-99-4b}]{\begin{minipage}{0.25\columnwidth}\input{variantsGo/handicap-99-4b}\end{minipage}}
  \hfill
  \subfloat[\label{fig:handicap-99-4c}]{\begin{minipage}{0.25\columnwidth}\input{variantsGo/handicap-99-4c}\end{minipage}}
  \hfill
  \subfloat[\label{fig:handicap-99-4d}]{\begin{minipage}{0.25\columnwidth}\input{variantsGo/handicap-99-4d}\end{minipage}}
  \hfill
  \subfloat[\label{fig:handicap-99-4e}]{\begin{minipage}{0.25\columnwidth}\input{variantsGo/handicap-99-4e}\end{minipage}}
  \hfill
  \subfloat[\label{fig:handicap-99-4f}]{\begin{minipage}{0.25\columnwidth}\input{variantsGo/handicap-99-4f}\end{minipage}}
  \hfill
  \subfloat[\label{fig:handicap-99-4g}]{\begin{minipage}{0.25\columnwidth}\input{variantsGo/handicap-99-4g}\end{minipage}}
  \hfill
  \subfloat[\label{fig:handicap-99-4h}]{\begin{minipage}{0.25\columnwidth}\input{variantsGo/handicap-99-4h}\end{minipage}}
  \hfill
  \subfloat[\label{fig:handicap-99-4i}]{\begin{minipage}{0.25\columnwidth}\input{variantsGo/handicap-99-4i}\end{minipage}}
  \hfill
  \subfloat[\label{fig:handicap-99-4j}]{\begin{minipage}{0.25\columnwidth}\input{variantsGo/handicap-99-4j}\end{minipage}}
  \caption{\small \label{fig:handicap-9x9} Handicap positions used for $9\times 9$ KAG.}
\end{figure}\end{small}}
{\begin{small}\begin{figure}[t]\centering\myfs
  \centering
  \hfill
  \subfloat[\label{fig:handicap-13-10a}]{\begin{minipage}{43mm}\input{variantsGo/handicap10a}\end{minipage}}
  \hfill
  \subfloat[\label{fig:handicap-13-10b}]{\begin{minipage}{43mm}\input{variantsGo/handicap10b}\end{minipage}}
  \hfill
  \subfloat[\label{fig:handicap-13-10c}]{\begin{minipage}{43mm}\input{variantsGo/handicap10c}\end{minipage}}
  \hfill
  \subfloat[\label{fig:handicap-13-10d}]{\begin{minipage}{43mm}\input{variantsGo/handicap10d}\end{minipage}}
  \hfill
  \caption{\small \label{fig:handicap-13x13}
    Various 10 stones handicaps on size $13 \times 13$.
    H10 refers to 10 stones in all the 3,7 points; the 3,3 points; and 6,7 and 8,7.}
\end{figure}\end{small}}
}{}
\def\uselessmaybe{If the winning rate of the weakest player is decreased in a variant $v$ (e.g. with PR), then, assuming that this effect holds for all players, we can guess that the Elo difference is increased in $v$, and therefore the game $v$ is deeper. 
Note that when comparing two distinct games with the same set of players, nothing ensures that being stronger for the original game is equivalent to being stronger for $v$. However, we will assume that when the games are very close (such as Go and KAG) the comparison is relevant.
Thus, we use the winrate to compare the {\em{depth}} of various KAG handicap placements.
This allows to estimate which starting position is fairer which in turn enables estimating how the {\em{depth}} evolves when the PR is applied.
Notwithstanding the caveat mentionned above, we also obtain a guess as to whether KAG with the PR is deeper than standard Go.
Fix a pair of players, and call them, say \emph{Weaky} and \emph{Strongy},
and define a set of a possible handicap placements $\{1, 2, \dots, i, \dots\}$. This setting gives birth to various winrates.
\def\racontons{Let $p_i$ be the winrate of Weaky against Strongy on position $i$ when Weaky is playing the Killer role, and let $q_i$ be the winrate of Weaky against Strongy on the same position but with the inversed roles.
$p_i$ and $q_i$ are determined experimentally, but we use them to define the expected winrate of Weaky against Strongy on position $i$ as $\frac{p_i+q_i}{2}$ if the roles are picked at random.
We also provide winrate on standard Go, that is non KAG and without handicap; $s$ is Weaky's winrate when playing Black and $s'$ is the winrate when playing White.}
The relations between the various winrates are summarized in Table~\ref{tab:winrate} which provides a template for the actual data.
The actual experiments involved board sizes $7\times 7$, $9\times 9$, and $13 \times 13$.
The player pairs consisted of two instances of the same program with different parameter settings so as to have on player stronger than the other.
We included the GnuGo program in alpha-beta mode as well as an MCTS version of GnuGo, plus ColdMilk and HappyGo, two MCTS programs.
\begin{table}[t!]\small
  \centering
  \caption{\small \label{tab:winrate}Organization of the winrates in a table.}
 \subfloat[]{ \begin{tabular}{lrrr}
    \toprule
             & As Killer & As Defender& Average\\
    \midrule
    $1$    &     $p_1$ & $q_1$ & $\frac{p_1 + q_1}{2}$\\
    $2$    &     $p_2$ & $q_2$ & $\frac{p_2 + q_2}{2}$\\
    \dots    &     \dots & \dots  & \\
    $i$    &     $p_i$ & $q_i$ & $\frac{p_i + q_i}{2}$\\
    \dots    &     \dots & \dots  & \\
    \cmidrule{1-4}
    RDR+PR & \multicolumn{3}{c}{$\frac12\left( \max_i \min(p_i,q_i)+ \min_i \max(p_i,q_i) \right)$}\\
       RDR & $\max_i p_i$ & $\min_i q_i$ & $\frac12(\max_i p_i+\min_i q_i)$ \\
    \midrule
    Standard-Go &       $s$ & $s'$   & $\frac{s+s'}{2}$ \\
    \bottomrule
  \end{tabular}}\hspace*{2mm}
\subfloat[]{
\begin{minipage}{5cm}
\racontons
\end{minipage}
}
\end{table}
Fig. \ref{kgo99} (right) shows that for a larger set of players (i.e. not in the framwork studied in the theorem above) and for choosing the number of handicap stones, PR is quite effective, by adapting the challenge to the level of players.

\subsubsection{Selecting handicap placement by {\em{depth}} analysis}\label{dkgsubsection}
We have seen that fixing the initial position (RDR-GFM$_i$) might lead to a deeper game than the PR (RDR+PR) when the first move is chosen so as to maximize the depth. The first kind of information that we can extract from these tables 
(Tables \ref{tab:77refinments} and \ref{tab:99refinments}) 
is which placement seem to discriminate most between a weak player and a stronger player (i.e., maximum \ac{PLC}, i.e. maximum Elo difference between these two players).
It appears that the deepest position among the proposed handicap placement on $7\times 7$ KAG is \ifthenelse{\longversion=1}{Fig.~\ref{fig:handicap-77-3g}. }{position (g) \cite{longversion}.}
Similarly, the deepest position among the proposed initial $9\times 9$ KAG positions seems to be \ifthenelse{\longversion=1}{Fig.~\ref{fig:handicap-99-4h}.}{position (h) \cite{longversion}.}
Indeed, the results show experimentally that the effect of the PR vary significantly with the set of positions to choose from.
On size $7\times 7$, if we restrict the players to the first 4 placements, the RDR+PR winrate is $42.2\%$, but if we add 8 more possible placements, the RDR+PR winrate rises to $44.9\%$.
This is a non-contrived example that a game with few options at the start might be deeper than a game with a strict superset of options.
The phenomenon does not appear on size $9\times 9$ through these experiments.
}
}
\def\boncestobscurcetruc
{
On the one hand, \ifthenelse{\longversion=1}{Fig.~\ref{fig:handicap-77-3g}}{position (7x7g) \cite{longversion}} seemingly makes $7\times7$ KAG deeper than standard $7\times 7$ Go.
On the other hand, \ifthenelse{\longversion=1}{Fig.~\ref{fig:handicap-99-4h}}{position (7x7-h) \cite{longversion}} does not seem to make $9\times 9$ KAG deeper than standard $9\times 9$ Go.
As for KAG extended with the PR, the RDR+PR numbers show that $7\times 7$ KAG is slightly deeper than standard $7\times 7$ Go, but $9\times 9$ KAG is significantly shallower than standard $9\times 9$ Go.
Eqs. \ref{plcelo} and \ref{iAmInAPeacefulPlace} are theoretically equivalent when the Elo model is valid. However, all results show they are different and that Eq. \ref{iAmInAPeacefulPlace} is more relevant.
We believe that Eq.~\ref{iAmInAPeacefulPlace} (based on a set of players) is a better tool for estimating the depth, and the intuitive notion of complexity, than the direct Elo difference between the strongest and the weakest player. 
\begin{table}[t!]\small
  \centering
  \caption{\small
    Refinments with various boards for size $7\times 7$ with HappyGo 400 against HappyGo 500.
    For these players, the PR makes KAG deeper, and indeed deeper than 7x7 Go, only when 4 handicap positionings are proposed; increasing the number of possible initial moves can make the game shallower.}
  \label{tab:77refinments} 
  \subfloat[Results for a set of 4 boards.]{
    \label{tab:77-4boards}
    \begin{tabular}{lrrr}
      \toprule
                                    & Killer & Defender & Average \\
      \midrule
      Fig.~\ifthenelse{\longversion=1}{\ref{fig:handicap-77-3a}}{7x7-a in \cite{longversion}} & $55.6\%$ & $36.6\%$ & $46.1\%$ \\
      Fig.~\ifthenelse{\longversion=1}{\ref{fig:handicap-77-3b}}{7x7-b in \cite{longversion}} & $54.6\%$ & $34.2\%$ & $44.4\%$ \\
      Fig.~\ifthenelse{\longversion=1}{\ref{fig:handicap-77-3c}}{7x7-c in \cite{longversion}} & $59.6\%$ & $29.4\%$ & $44.5\%$ \\
      Fig.~\ifthenelse{\longversion=1}{\ref{fig:handicap-77-3d}}{7x7-d in \cite{longversion}} & $38.8\%$ & $45.6\%$ & $42.2\%$ \\
      \midrule
      {\bf{RDR+PR}}                           & \multicolumn{3}{c}{$42.2\%$} \\
      RDR                           & $59.6\%$ & $29.4\%$ & $44.5\%$ \\
      \midrule
      Standard-Go                   & $36.2\%$ & $50.8\%$ & $43.5\%$ \\
      \bottomrule
    \end{tabular}
  }
  \subfloat[Results for a superset of 12 boards.]{
    \label{tab:77-12boards}
    \begin{tabular}{lrrr}
      \toprule
                                    & Killer & Defender & Average \\
      \midrule
      Fig.~\ifthenelse{\longversion=1}{\ref{fig:handicap-77-3a} to \ref{fig:handicap-77-3d}}{a to d\cite{longversion}} & \multicolumn{3}{c}{See Table~\subref{tab:77-4boards}}\\
      Fig.~\ifthenelse{\longversion=1}{\ref{fig:handicap-77-3e}}{7x7-e\cite{longversion}} & $37.4\%$ & $53.4\%$ & $45.4\%$ \\
      Fig.~\ifthenelse{\longversion=1}{\ref{fig:handicap-77-3f}}{7x7-l\cite{longversion}} & $47.2\%$ & $44.2\%$ & $45.7\%$ \\
      Fig.~\ifthenelse{\longversion=1}{\ref{fig:handicap-77-3g}}{7x7-l\cite{longversion}} & $48.0\%$ & $34.8\%$ & $41.4\%$ \\
      Fig.~\ifthenelse{\longversion=1}{\ref{fig:handicap-77-3h}}{7x7-l\cite{longversion}} & $57.4\%$ & $30.2\%$ & $43.8\%$ \\
      Fig.~\ifthenelse{\longversion=1}{\ref{fig:handicap-77-3i}}{7x7-l\cite{longversion}} & $32.2\%$ & $51.4\%$ & $41.8\%$ \\
      Fig.~\ifthenelse{\longversion=1}{\ref{fig:handicap-77-3j}}{7x7-l\cite{longversion}} & $32.2\%$ & $59.2\%$ & $45.7\%$ \\
      Fig.~\ifthenelse{\longversion=1}{\ref{fig:handicap-77-3k}}{7x7-l\cite{longversion}} & $41.4\%$ & $49.2\%$ & $45.3\%$ \\
      Fig.~\ifthenelse{\longversion=1}{\ref{fig:handicap-77-3l}}{7x7-l\cite{longversion}} & $43.8\%$ & $52.0\%$ & $47.9\%$ \\
      \midrule
      RDR+PR                           & \multicolumn{3}{c}{$44.9\%$} \\
      {\bf{RDR}} & $57.4\%$ & $30.2\%$ & $43.8\%$ \\
      Standard-Go                      & $36.2\%$ & $50.8\%$ & $43.5\%$ \\
      \bottomrule
    \end{tabular}
  }
\end{table}
\begin{table}[t!]\small
  \centering
  \caption{\small \label{tab:99refinments} 
    Refinments with various boards for size $9\times 9$ with HappyGo 400 against HappyGo 500.  In this case PR does not improve the depth.}
  \subfloat[Results for a set of 5 boards.]{
    \label{tab:99-5boards} 
    \begin{tabular}{lrrr}
      \toprule
                                    & Killer & Defender & Average \\
      \midrule
      Fig.~\ifthenelse{\longversion=1}{\ref{fig:handicap-99-4a}}{9x9-a \cite{longversion}} & $53.6\%$ & $37.2\%$ & $45.4\%$ \\
      Fig.~\ifthenelse{\longversion=1}{\ref{fig:handicap-99-4b}}{9x9-b\cite{longversion}} & $52.8\%$ & $41.6\%$ & $47.2\%$ \\
      Fig.~\ifthenelse{\longversion=1}{\ref{fig:handicap-99-4c}}{9x9-c\cite{longversion}} & $48.8\%$ & $47.4\%$ & $48.1\%$ \\
      Fig.~\ifthenelse{\longversion=1}{\ref{fig:handicap-99-4d}}{9x9-d\cite{longversion}} & $49.2\%$ & $42.2\%$ & $45.7\%$ \\
      Fig.~\ifthenelse{\longversion=1}{\ref{fig:handicap-99-4e}}{9x9-e\cite{longversion}} & $52.6\%$ & $38.2\%$ & $45.4\%$ \\
      \midrule
      RDR+PR                           & \multicolumn{3}{c}{$48.1\%$} \\
      {\bf{RDR}}          & $53.6$ & $37.2\%$ & $45.4 \%$ \\
      \midrule
      Standard                      & $35.4\%$ & $43.6\%$ & $39.5\%$ \\
      \bottomrule
    \end{tabular}
  }
  \subfloat[Results for a set of 10 boards.]{
    \begin{tabular}{lrrr}
      \toprule
                                    & Killer & Defender & Average \\
      \midrule
      Fig.~\ifthenelse{\longversion=1}{\ref{fig:handicap-99-4a} to \ref{fig:handicap-99-4e}}{9x9-a--e \cite{longversion}} & \multicolumn{3}{c}{See Table~\subref{tab:99-5boards}}\\
      Fig.~\ifthenelse{\longversion=1}{\ref{fig:handicap-99-4f}}{9x9-f \cite{longversion}} & $51.0\%$ & $42.0\%$ & $46.5\%$ \\
      Fig.~\ifthenelse{\longversion=1}{\ref{fig:handicap-99-4g}}{9x9-g \cite{longversion}} & $52.4\%$ & $41.6\%$ & $47.0\%$ \\
      Fig.~\ifthenelse{\longversion=1}{\ref{fig:handicap-99-4h}}{9x9-h \cite{longversion}} & $48.8\%$ & $40.6\%$ & $44.7\%$ \\
      Fig.~\ifthenelse{\longversion=1}{\ref{fig:handicap-99-4i}}{9x9-i \cite{longversion}} & $53.4\%$ & $38.4\%$ & $45.9\%$ \\
      Fig.~\ifthenelse{\longversion=1}{\ref{fig:handicap-99-4j}}{9x9-j \cite{longversion}} & $48.4\%$ & $43.0\%$ & $45.7\%$ \\
      \midrule
      RDR+PR                           & \multicolumn{3}{c}{$47.9\%$} \\
											  {\bf{RDR}}  & $53.4\%$ & $38.4\%$ & $45.9\%$ \\
      \bottomrule
    \end{tabular}
  }
\end{table}
We realized three sets of experiments on size $13 \times 13$.
We report results for GnuGo level 1 against level 10,
ColdMilk with 400 and 500 simulations per move,
and ColdMilk with 3000 simulations per move against 5000 simulations per move in Table~\ref{tab:kgoxps}.
The results of the tests involving GnuGo suggest that $13\times 13$ KAG with 8-stone handicap and $13\times 13$ KAG with 9-stones handicap are both deeper than $13\times 13$ Standard Go; the difference is significant for 8-stone handicap. 
It has sometimes been pointed out that {\em{depth}} should be normalized by the length of games --- but indeed the length of games was \emph{smaller} for KAG.
Therefore, even after such a normalization, these results would seem to indicate that handicap 8 $13\times 13$ KAG is deeper.
As for the PR, it seems to lead to a game roughly as deep as standard Go.
The results involving ColdMilk also indicate that the game extended with the PR is as deep as standard Go.
On the other hand, contrary to the results with GnuGo, the ones with ColdMilk do not point out any of the single handicap stones placements as being significantly deeper than standard Go.
The first player chooses the handicap, and indeed always chooses H10 in RDR; so the Pie rule has a positive effect.
The detailed results suggest however that H10 is too easy for Black for Coldmilk and H9 is too easy for White.
It is not surprising that the relevant number of handicap stones for making the game fair highly depends on the strength of the players and their playing style. }
\section{Conclusions}\label{conc}
{\bf{Game complexity measures: beyond computational complexity.}} Depth and \ac{PLC} have fundamental shortcomings. They depend on a set of players. Maybe the dependency on a set of players is necessary, in the sense that, as we are looking for a measure relevant for humans, we need humans or human-like AIs somewhere in the definition. One can also easily construct trivial games (e.g. who has the longest right foot) which are very deep. 
Nonetheless {\em{depth}} is interesting as a measure for comparing games. \ac{PLC} is a nice refinement, and the best formula for measuing it is Eq. \ref{iAmInAPeacefulPlace} as discussed below. 

{\bf{Elo model.}} A side conclusion is that the Elo model, at least when pushed to the limit for such \ac{PLC} analysis, is not verified (see also \cite{fuckelo}). Under the Elo model, Eq. \ref{iAmInAPeacefulPlace} and Eq. \ref{plcelo} should be equivalent, which is not the case in our experiments. {\color{black}We see that, consistently with \cite{fuckelo}, the Elo model poorly modelizes large rank differences. For example, IRT models\cite{irt} include lower and upper bounds ($>0$ and $<1$ respectively) on the probability of succeeding, regardless of the ranks; this might be used in Elo scales as well. Further analysis of this point is left for further work.}

{\bf{The pie rule and the \ac{PLC}.}}
 The PR does not necessarily preserve the \ac{PLC} of a game, compared to a fixed good choice of the initial move. In fact, with 2 players only in the considered set, it cannot increase the \ac{PLC} compared to the variant of the game which enforces the first move to the deepest opening move (i.e. if it enforces the first move to the one which leads to the deepest game for these two players).
However, it sometimes improves the \ac{PLC}, with more than 2 players, or compared to a poor choice (in the rules) of an initial move, or compared to the classical case of a choice of the initial move by the first player. 
PR is very effective when very strong first moves exist. Theorem \ref{thm:pr-bounded} does not contradict this, as it discusses the case in which the initial move is chosen by the referee. Basically PR discards such first moves from the game, so that the referee does not have to do it. Even with just two players, PR can increase the depth, compared to RDR, which is the most usual case. Our counter-example can be extended to several players, but then it only shows that the pie rule does {\em{not always}} increase the depth; whereas for two players we have the stronger result that the pie rule {\em{cannot}} increase the depth compared to an ad hoc choice of the first move - i.e. there is at least one first move such that, if it is imposed by the rules, the Elo difference between these two players will be at least the same as with pie rule.

 {\bf{Experimental results on various real-world games: do we improve the {\em{depth}} when using the pie rule ?}}
For the game of Y, tested with players with strongly varying strength and moderate board size, the PR is very effective, making the game deeper than any other rule for choosing the first move.
Concerning the NoGo game, even if PR made the game deeper compared to RDR, an ad hoc choice of the first move by an omniscient referee leads to the deepest variant of the game.
In KAG, the pie-rule is clearly successful for choosing the number of handicap stones in KAG. The wider the range of considered players, the better the PR (Fig. \ref{kgo99}, right). 

{\bf{Analyzing games with {\em{depth}} analysis.}}
Depth and \ac{PLC} are also tools for quantifying (for sure not in a perfect manner) the importance and challenging nature of a game. 
Interestingly, KAG is deeper than Go, by very far for some set of players we have considered, namely variants of GnuGo, in 9x9 - whereas Go is usually considered as the deepest known game. This is however unstable; there are some thresholds at which the winning rate between $i$ and $i+50$ simulations are very large (e.g 550 simulations per move vs 500 simulations per move), and when such gaps are excluded from the set of players, the depth of KAG does not increase that much with pie rule. 

{\bf{Depth as a criterion for choosing exercises.}} KAG is widely used as an exercise and is particularly deep; this suggests that {\em{depth}} might be a good criterion for choosing problems. Generating problems is not that easy, in particular in games such as Go - depth is a criterion for selecting interesting positions.

{\bf{Chinese Dark Chess.}} Just doubling the width of CDC makes the game much deeper and far more challenging (Table \ref{tab:cdc}). All rules are preserved. This might give birth to a new game for humans.

{\bf{The first move in NoGo.}} Last, experiments on NoGo show that the opening move is often not the rational one and that the opening move has a big impact on the game (see Eq. \ref{nogo1}-\ref{nogo3}, showing the depth when using a real first move, compared to Table \ref{tab:nogo-y} which uses a rational first move).
\bibliographystyle{IEEEtran}
\bibliography{test}
\end{document}